\documentstyle[aps,prl,twocolumn,epsf]{revtex}
\newcommand{\beq}{\begin{equation}}
\newcommand{\eeq}{\end{equation}}
\newcommand{\bea}{\begin{eqnarray}}
\newcommand{\eea}{\end{eqnarray}}

\begin{document}

\title{Selfconsistent effective interactions and symmetry restoration }

\author{ Piotr MAGIERSKI\thanks{On
leave of absence from Institute of Physics,
Warsaw University of Technology, Warsaw, Poland.} and Ramon WYSS }

\address{$^1$ The Royal Institute of Technology, Physics Department Frescati,
Frescativ\"agen 24, S--10405, Stockholm, SWEDEN }

\date{\today }

\maketitle

\begin{abstract}

In this  paper we propose a general framework for deriving the effective
interactions for  many-body systems. 
We show how the selfconsistent interaction can be constructed on the quantum
level for both local and nonlocal potentials.
In particular the relation between the selfconsistent and symmetry restoring interaction
is derived. The difference between the two is related 
to the coupling of the perturbing field to the kinetic energy operator. 

\end{abstract}


\vspace{0.5cm}

\narrowtext

For the study of collective motion of the nuclear or other many-body
systems it is essential to properly define the residual interactions.
This problem appears always when the ground
state has been obtained by means of some variational principle without the
explicit use of many-body forces.
Based on the requirement of selfconsistency, Bohr and Mottelson
derived the effective interactions
for spherical nuclei\cite{BM}.
For a deformed system, like the
Nilsson potential, Marshalek has shown
that the volume conservation constraint is equivalent to a
Hartree approximation for many-body forces \cite{Mar}. 
The effective interaction can be formally constructed as 
functional derivatives of the total energy
with respect to the density. For the harmonic oscillator potential
it can be done analytically \cite{Kis1,Sak1}.
Based on this method and the Thomas-Fermi approximation
it has been shown how the subsequent terms of the 
effective interaction can be obtained \cite{Kis2,Sak2}.

However, all these methods unnecessarily
rely on semiclassical approximations. 
Moreover, the comparison between the selfconsistent strengths
of the interaction and the so-called microscopic ones based on the linear response
theory can result in different values \cite{Sak1}.
The nature of this difference, albeit attributed to shell effects, 
seemed not to be clear, but can be elucidated on the quantum level.
Last not least, there is a powerful symmetry restoration method introduced
originally by Pyatov and collaborators \cite{Pya1} and developed in a series
of papers \cite{rest},
which appears to be related to the concept of selfconsistent forces. 
To our knowledge 
this link has not been established yet.
In the present paper we show how
one can formulate the problem on the quantum level.
We will not address any specific choice of coordinates which could enable
us to decouple collective modes from  spurious
components. This can be done at any stage of the presented method, although it
is not crucial for the merit of this article.

Let us assume that our phenomenological one body hamiltonian has the form:
\begin{equation} \label{ham}
\hat{H}=\hat{T}+\hat{V},
\end{equation}
with the kinetic energy
$\hat{T}=\displaystyle{\frac{\hat{\bf p}^2}{2m}}$ and the potential 
$\hat{V}( {\bf r}, {\bf r}' )$.
The nonlocality  of the potential $\hat{V}$ can appear e.g. as
a consequence of a spin-orbit coupling. 
The solution of the hamiltonian (\ref{ham}) defines a one body density 
$\rho ({\bf r},{\bf r'})=
\displaystyle{\sum_{k=1}^{N}}\phi_{k}^{*}({\bf r})\phi_{k}({\bf r'})$,
where $\phi_{k}({\bf r})$, $k=1,...,N$ denote the  $N$ lowest eigenfunctions 
of $\hat{H}$, and the density operator is defined via
$\hat{\rho}=\displaystyle{\sum_{k=1}^{N}} | k \rangle\langle k|$.
To lowest order the hamiltonian describing the selfconsistent interaction
is assumed to have the separable form:
\begin{equation} \label{inter}
\hat{H}_{int}=\frac{1}{2}\chi \hat{Q}^{2},
\end{equation}
where $\hat{Q}$ is a hermitian one-body operator and $\chi$ denotes a coupling constant. 
The reason that the particular form of the interaction
(\ref{inter}) is used, 
instead of the more general: $\hat{Q'}^{+}\hat{Q'}$, with $\hat{Q'}^{+}\neq \hat{Q'}$,
is associated with the fact that the exchange potential 
generated by the separable interaction can be neglected. Hence
the Hartree fields of the interaction $\hat{Q'}^{+}\hat{Q'}$ and (\ref{inter}) 
are equivalent if $\hat{Q}$ is a hermitian or antihermitian part of $\hat{Q'}$.
Also in the RPA approach the interaction $\hat{Q'}^{+}\hat{Q'}$ can always
be transformed to the form (\ref{inter}) in the quasiboson approximation.

The following conditions should be imposed: (i)
$ Tr(\hat{Q}\hat{\rho})=\langle\hat{Q}\rangle_{0}=0$,
(ii) the interaction $\hat{H}_{int} $ should induce the same change
in the density $\rho$ as in the potential $\hat{V}$.
Clearly, condition (i) provides a restriction on $\hat{Q}$,
whereas (ii)
determines the coupling constant.
Condition (i) is rather difficult to fulfill for an arbitrary potential,
except when $\hat{V}$ is spherical. For the deformed harmonic oscillator,
Sakamoto and Kishimoto derived the 
double-stretched
multipole interaction that always
fulfills criterion (i)\cite{Sak1}.
This interaction has the further advantage to exactly decouple the
spurious $K=1$ modes from the shape multipole oscillations.
Recently Kubo et al. \cite{Kubo} extended these methods to
momentum dependent potentials.
For realistic potentials like the Woods-Saxon potential, condition
(i) can be fullfilled approximately in a similar manner \cite{Sat93}.

To analize the role of the condition (ii),
assume that the collective mode is produced by the
one body field $\hat{A}$. Hence, the operator $\hat{U}=\exp (-i \hat{A} \theta)$
defines the transformation of the ground state ($N-$body) wave function:
\begin{equation}
U^{+}|\psi\rangle \approx |\psi\rangle + i \theta \hat{A} |\psi\rangle + ...
\end{equation}
Since we limit ourselves to small perturbations we will consider
only the first term of the expansion. 
In order to preserve the normalization of the wave function we require
$\hat{A}$ to be hermitian ($\theta$ is real). 
Instead of changing the wave function it is easier to consider the 
change of the operators $\hat{V}$ and $\hat{\rho}$:
\begin{eqnarray} \label{trans}
U^{+}\hat{V}U & \approx & \hat{V} + i \theta [\hat{A},\hat{V}]  \\
U^{+}\hat{\rho }U & \approx & \hat{\rho } + i \theta [\hat{A},\hat{\rho } ] .
\end{eqnarray}
The condition (ii) requires that $\theta$ is proportional  to 
$\langle [\hat{A},\hat{V}]\rangle$:
\begin{equation}
\theta=i\chi\langle [\hat{A},\hat{V}] 
\rangle=i\chi Tr( [\hat{A},\hat{V}] U^{+}\hat{\rho }U ), 
\end{equation}
which leads to the equation:
\begin{equation} \label{phi}
\theta=i\chi\langle [\hat{A},\hat{V}] \rangle_{0}-\chi\theta\langle 
[ [\hat{A},\hat{V}],\hat{A} ] \rangle_{0}. 
\end{equation}
Since in general the first term need not to vanish, an arbitrary 
field $\hat{A}$
does not generate a selfconsistent mode unless one will make the replacement:
$[\hat{A},\hat{V}] \rightarrow [\hat{A},\hat{V}]-\langle [\hat{A},\hat{V}]\rangle_{0}$. 
Only when: $\langle [\hat{A},\hat{V}] \rangle_{0}=0$, 
eq.~(\ref{phi}) defines the selfconsistent interaction strength  $\chi^{self}$,
\begin{equation} \label{selcoup}
\frac{1}{\chi^{self}}=-\langle [ [\hat{A},\hat{V}],\hat{A} ] \rangle_{0}.
\end{equation}
Thus the selfconsistency requirement specifies the
interaction:
\begin{equation} \label{int1}
\hat{H}_{int}=\frac{1}{2}\chi^{self}\hat{Q}^2=-\frac{1}{2}\chi^{self} [\hat{A},\hat{V}]^2.
\end{equation}

Note, that if one chooses $\hat{A}$ as an operator associated with a particular
symmetry like e.g. angular momentum component $\hat{J}_{i}$, 
or total momentum $\hat{P}$, then 
the above prescription is equivalent to the Pyatov formula for the symmetry
restoring interaction\cite{Pya1}. Since these operators commute with 
the kinetic energy
$\hat{T}$,
one can write (\ref{int1}) in the form:
\begin{equation} \label{int2}
\hat{H}_{int}=-\frac{1}{2}\chi^{rest} [\hat{A},\hat{H}]^2, 
\end{equation}
with the symmetry restoring coupling strength:
\begin{equation} 
\frac{1}{\chi^{rest}}=-\langle [ [\hat{A},\hat{H}],\hat{A} ] \rangle_{0}.
\end{equation}

In order to consider excitations that cause a change of the density,
we need to find an explicit form of 
the operator $\hat{A}$. Starting from the
single-particle wave functions we note that they
will be transformed according to:
\begin{equation} \label{trwv}
\phi_{k}( {\bf r} ) \rightarrow 
\frac{1}{\sqrt{1-{\bf \nabla}\cdot{\bf u}}} \phi_{k}( {\bf r}+ {\bf u}({\bf r}) ),
\end{equation}
where ${\bf u}$ is in general a function of ${\bf r}$. 
Since the ground state
is an independent particle state, eq.~(\ref{trwv}) generates
a transformation of the many body wave function $\psi$.
The Pauli principle is not violated by this transformation 
because all the wave functions are transformed in the same 
manner \cite{ber1}.
The transformation (\ref{trwv}) is generated by an operator
 $\hat{A}$ of the form:
\begin{equation} 
\hat{A}=\frac{1}{2}(\hat{\bf u}({\bf r})\cdot\hat{\bf p}+
                    \hat{\bf p}\cdot \hat{\bf u}({\bf r}) ),
\end{equation}
and
\begin{equation}
U^{+}\phi_{k}( {\bf r} ) \approx  \phi_{k}( {\bf r} ) + 
\theta {\bf u} \cdot {\bf \nabla}\phi_{k}( {\bf r} ) +
\frac{1}{2}\theta \phi_{k}( {\bf r} ){\bf \nabla}\cdot{\bf u}.  
\end{equation}

The field $\hat{A}$ contains e.g. all the $2^{\lambda}$-pole 
distortions. Indeed, by choosing 
${\bf u}({\bf r})={\bf \nabla} \tilde{Q}_{\lambda\mu}$,
where 
$\tilde{Q}_{\lambda\mu}=\displaystyle{\frac{1}{\sqrt{2}}}(Q_{\lambda\mu}+Q^{+}_{\lambda\mu})$
or 
$\tilde{Q}_{\lambda\mu}=\displaystyle{\frac{i}{\sqrt{2}}}( Q_{\lambda\mu}-Q^{+}_{\lambda\mu})$,
the wave function $\psi$ will be affected by the multipole field.

Following the above prescription one can easily (at least formally) construct
the appropriate  selfconsistent interaction:
\begin{eqnarray} \label{int}
\hat{H}_{int}&=&-\frac{1}{2}\chi^{self}
[ \hat{\bf u}\cdot\hat{\bf p}+\hat{\bf p}\cdot\hat{\bf u}, \hat{V}]^2, \\
\frac{1}{\chi^{self}}&=&
-\langle [ [ \hat{\bf u}\cdot\hat{\bf p}+\hat{\bf p}\cdot\hat{\bf u}, \hat{V}],
      \hat{\bf u}\cdot\hat{\bf p}+\hat{\bf p}\cdot\hat{\bf u} ]\rangle_{0}. 
\nonumber
\end{eqnarray}
This interaction is not anymore limited to the volume conserving modes, 
${\bf \nabla}\cdot{\bf u}=0$.
Other excitations like the breathing mode can also be considered.
In the case of a multipole field and a potential which does not depend
on the momentum the relations (\ref{int}) reduces
to the expression:
\begin{eqnarray} \label{local}
\hat{H}_{int}&=&\frac{1}{2}\chi^{self} ( {\bf \nabla}\tilde{Q}_{\lambda\mu}\cdot{\bf \nabla} V )^2, \\
\frac{1}{\chi^{self}}&=&-\langle  
{\bf \nabla} ( {\bf \nabla}\tilde{Q}_{\lambda\mu}\cdot{\bf \nabla} V )\cdot
                          {\bf \nabla}\tilde{Q}_{\lambda\mu} \rangle_{0} .
\end{eqnarray}
This result differs slightly from that in Ref. \cite{Kubo} because we consider only a single
excitation mode. 

In the case of nonlocal potentials expression (\ref{local}) is not valid
since $[ {\bf \nabla}\hat{Q}_{\lambda\mu},\hat{V} ]\neq 0$ and one has to
use the general expression (\ref{int}). 
Note, that the transformation (\ref{trwv}) induces a deformation of the
Fermi sphere resulting in a change of the current distribution:
\begin{equation}
{\bf j}={\bf j}_{0}+({\bf u}\cdot{\bf \nabla} ){\bf j}_{0}+
({\bf j}_{0}\cdot{\bf \nabla} ){\bf u}+{\bf j}_{0}{\bf \nabla}\cdot{\bf u},
\end{equation}
where ${\bf j}=\displaystyle{\frac{1}{2i}\sum_{k=1}^{N}}
( \phi_{k}^{*}{\bf \nabla}\phi_{k}-({\bf \nabla}\phi_{k}^{*})\phi_{k} )$, and ${\bf j}_{0}$
denotes the equilibrium distribution.
Hence, an additional
correction arises for nonlocal
potentials.
This effect was not considered in previous approaches
since it vanishes in the hydrodynamical limit.
It should be mentioned that since the nonlocal potentials violate the
Galilean invariance its presence lead to the appearance of an
additional interaction discussed in Ref.\cite{Sak3}.

Let us now examine in more detail the concept of selfconsistent
and symmetry restoring interactions. If
the hamiltonian of eq.~(1) violates a given symmetry
one may restore it in the RPA approximation by adding 
an interaction of type (10).
The interaction strength
$\chi^{rest}$ given by the expression (11) assures that there are no restoring forces towards
the perturbation generated by the field $\hat{A}$ \cite{Pya1}.
For the harmonic oscillator model
the equality between the latter and the 
selfconsistent strength
holds only for a few particular cases whereas usually it is different 
by the value of  a discrete jump, attributed
to the influence of the shell effects \cite{Sak1}.
In the following we argue instead, that these jumps are related
to two different levels of symmetry restoration.
The selfconsistent interaction arises from the requirement
of the invariance of the potential towards the perturbation generated by field
$\hat{A}$, whereas the symmetry restoring forces assure the invariance of
the total hamiltonian.

As discussed above,
the operator $\hat{A}$ does not commute in general 
with the kinetic energy operator. Hence, one cannot replace the 
potential $\hat{V}$
by the total hamiltonian $\hat{H}$ in formula (\ref{selcoup}). 
This is a consequence of the fact that in general
the perturbed hamiltonian and density do not commute.
Namely:
\begin{equation}
[ \hat{H}( \rho+ \delta\rho ),
   \hat{\rho}+\delta\hat{\rho} ] \neq 0
\end{equation}
where $\hat{H}$ is a one body field generated by the density $\rho$ through
the Hartree(-Fock) procedure i.e. $[\hat{H}(\rho ),\hat{\rho}] = 0$.

This can easily be verified in the case of the harmonic oscillator model.
Consider the three-dimensional spherical harmonic oscillator hamiltonian:
\begin{equation}
\hat{H}=\frac{\hat{\bf p}^2}{2m}+\frac{m}{2}\omega ^2 {\bf r}^{2}.
\end{equation}
Assume that the operator $\hat{A}$ is choosen in the form:
\begin{equation}
\hat{A}={\bf \nabla} \hat{Q}_{20}\cdot \hat{\bf p}.
\end{equation}
Then the perturbed hamiltonian has the form:
\begin{equation}
\hat{H'}=\hat{H}+2m\omega^{2}\theta \hat{Q}_{20}({\bf r}).
\end{equation}
If we define $\alpha=4\theta$ then:
\begin{eqnarray}
\hat{H'}&=&\frac{\hat{\bf p}^2}{2m}+
\frac{m}{2}\omega ^2 
( (1-\alpha )x_{1}^{2}+(1-\alpha )x_{2}^{2}+(1+2\alpha )x_{3}^{2} )= \nonumber \\
& &\frac{\hat{\bf p}^2}{2m}+
\frac{m}{2}\sum_{i=1,2,3}\omega_{i} ^2 x_{i}^{2}.
\end{eqnarray}
For the harmonic oscillator the selfconsistency requirement between
the potential and the density takes the simple form:
\begin{equation} \label{self1}
\omega_{i}\Sigma_{i}=const.,
\end{equation}
where $\Sigma_{i}=m\omega_{i}Tr( x_{i}^{2}\hat{\rho} )$.
However,
if one calculates the eigenfunction of the perturbed hamiltonian
then relation (\ref{self1}) is violated, since
\begin{equation} \label{viol}
\omega_{i}\Sigma_{i}=( \omega +\delta\omega_{i} )\Sigma_{i},
\end{equation}
where 
\begin{eqnarray}
\delta\omega_{1}&=&\delta\omega_{2}=-\frac{1}{2}\alpha\omega + O(\alpha^{2}) \nonumber 
\\
\delta\omega_{3}&=&\alpha\omega + O(\alpha^{2}). 
\end{eqnarray}
We assumed that there are no configuration changes
caused by the perturbation which obviously is the case for the
small perturbations considered in this article.

To preserve the selfconsistency in the perturbed system one has
to transform the wave function according to eq.~(3), where the
operator $\hat A$ is defined by eq.~(21). 
It defines the new single particle wave functions 
$\phi_{k}'({\bf r})=\phi_{k}({\bf r'})$, where
\begin{equation} 
{\bf r'} = {\bf  r} +\frac{\alpha}{4} {\bf \nabla} \hat{Q}_{20}.
\end{equation}
Using such  wave functions one can easily find that:
\begin{equation} 
\omega_{i}\Sigma_{i}=\omega \sum_{k=1}^{N} (n_{k}+\frac{1}{2} )=const. ,
\end{equation}
and the relation (\ref{self1}) is preserved.

Hence we see that two methods of obtaining the perturbed wave function do not 
coincide.
Only one of them generates the selfconsistent density of the perturbed system.
This difference is responsible for the difference between the selfconsistent
and symmetry restoring coupling constants.
To make the discussion more general we use now the response function formalism
to show when these two concepts lead to the same results.

Consider the perturbation of the hamiltonian defined as:
\begin{equation}
\delta\hat{H}=i[\hat{A},\hat{H}]=i[\hat{A},\hat{T}]+i[\hat{A},\hat{V}]=
\delta\hat{T}+\delta\hat{V}.
\end{equation}
The total hamiltonian has the form
\begin{equation}
\hat{H'}=\hat{H}+\delta\hat{H}\theta (t),
\end{equation}
where the perturbation in general can be time-dependent.
We define the quantity: 
\begin{equation}
\Delta E (t) = \langle\hat{H'}\rangle-\langle\hat{H'}\rangle_{0},
\end{equation}
which represent the difference between the expectation value of the hamiltonian
in the perturbed and initial state.
The Fourier transform of this quantity is related to the Fourier transform
of the function $\theta (t)$:
\begin{equation} \label{re}
\Delta E (\omega)= R_{\delta H, \delta H}(\omega )\theta(\omega ).
\end{equation}
The response function $R_{\delta H, \delta H}(\omega )$ has the form:
\begin{equation}
R_{\delta H, \delta H}(\omega )=
-\sum_{k>0}\frac{|\langle k|\delta\hat{H}|0\rangle |^{2} 
(E_{k}-E_{0})}{(E_{k}-E_{0})^{2}-\omega^{2}},
\end{equation}
where $E_{0}$ and $E_{k}$ denote the energies of the ground state $|0\rangle $
and excited states $|k\rangle $, respectively.

On the other hand the evaluation of the left-hand side of Eq. 
(\ref{re}) at $\omega=0$ gives:
\begin{equation}
\Delta E (\omega =0)=-\langle [[\hat{A},\hat{H}],\hat{A}]\rangle_{0}.
\end{equation}
Hence in the case when $[\hat{T},\hat{A}]=0$ 
we get:
\begin{equation}
\frac{1}{\chi^{self}}=R_{\delta H, \delta H}(\omega =0 )=R_{\delta V, \delta V}(\omega =0 ),
\end{equation}
where $\chi^{self}$ has been defined according to (\ref{selcoup}).
This result is exactly the prescription for determining the symmetry
restoring
coupling strength. 
One can see that it coincides with the 
selfconsistent strength 
when the operator $\hat{A}$ commutes with the kinetic energy operator.

In the case when $[\hat{T},\hat{A}]\neq 0$ one can rewrite
the equation (\ref{re}) in the form:
\begin{equation} \label{coupl}
\frac{1}{\chi^{self}}=R_{\delta V, \delta V}(\omega =0 )+
R_{\delta V, \delta T}(\omega =0 ),
\end{equation}
\begin{equation} 
R_{\delta V, \delta T}(\omega =0 )=-\sum_{k>0}
\Bigg{(}
\frac{\langle 0|\delta\hat{V}|k\rangle\langle k|\delta\hat{T}|0\rangle}{E_{k}-E_{0}}
+ c.c. \Bigg{)},
\end{equation}
where we have used the relation:
\begin{equation}
\langle [[\hat{A},\hat{T}],\hat{A}]\rangle_{0}=-R_{\delta T, \delta H}(\omega =0).
\end{equation}
Equation (\ref{coupl}) provides the connection between the microscopic
and selfconsistent coupling constants. Namely,
\begin{equation} 
\frac{1}{\chi^{self}}=\frac{1}{\chi^{rest}}+R_{\delta V, \delta T}(\omega =0 ).
\end{equation}

Hence, one can notice that the difference between the selfconsistent
and symmetry restoring
strength is related to the coupling of the field $\hat{A}$ to
the kinetic energy operator.  Whereas the selfconsistent strength has to be
calculated using the prescription  (\ref{selcoup}), 
the symmetry restoring one is defined as the
inverse of the response function at zero frequency and guarantees that there
must exist a spurious zero-energy RPA mode.

One can easily notice that $R_{\delta V, \delta T}(\omega =0)$ vanishes e.g. 
when
we consider dipole distortions. 
The operators $\hat{A}_{1 \mu}$ in this case has the form:
\begin{eqnarray} \label{q21}
\hat{A}_{10}&=&{\bf \nabla}Q_{10}\cdot {\bf p} \sim \hat{p}_{z} \nonumber \\
\hat{A}_{11}&=&{\bf \nabla}( Q_{11}-Q_{1-1} )\cdot {\bf p} \sim \hat{p}_{x} \\
\hat{A}_{\tilde{11}}&=&
i{\bf \nabla}( Q_{11}+Q_{1-1} )\cdot {\bf p} \sim \hat{p}_{y}. \nonumber 
\end{eqnarray}
This implies that:
\begin{equation} \label{tkin}
\delta\hat{T}=0.
\end{equation}
Consequently for the dipole mode the microscopic
and selfconsistent strength coincide: $\chi^{rest}_{1\mu}=\chi^{self}_{1\mu}$.

To conclude, in the present paper we have shown that the selfconsistent
interactions can be derived at the quantum level for general potentials. The Pyatov
prescription for the symmetry restoring forces can be included 
in this general framework.
The change of the momentum distribution will produce an additional
correction to the
selfconsistent interaction if momentum dependent potentials are considered.
The origin of the difference between the 
symmetry restoring
and selfconsistent coupling constants relates to two different levels of
symmetry restoration. Whenever the perturbing field commutes with the
kinetic energy operator the two concepts are equal.


\bigskip

PM acknowledge the financial support
from the Swedish Institute and the G\"oran Gustafsson foundation.

\end{document}